# Probing the Nucleation of $Al_2O_3$ in Atomic Layer Deposition on Aluminum for Ultrathin Tunneling Barriers in Josephson Junctions

Alan. J. Elliot[1], Gary Malek[1], Logan Wille[1], Rongtao Lu[1], Siyuan Han[1], Judy Z. Wu[1], John Talvacchio[2], Rupert M. Lewis[2].

*Abstract*— **Ultrathin dielectric tunneling barriers are critical to Josephson junction (JJ) based superconducting quantum bits (qubits). However, the prevailing technique of thermally oxidizing aluminum via oxygen diffusion produces problematic defects, such as oxygen vacancies, which are believed to be a primary source of the two-level fluctuators and contribute to the decoherence of the qubits. Development of alternative approaches for improved tunneling barriers becomes urgent and imperative. Atomic Layer Deposition (ALD) of aluminum oxide ($Al_2O_3$) is a promising alternative to resolve the issue of oxygen vacancies in the $Al_2O_3$ tunneling barrier, and its self-limiting growth mechanism provides atomic-scale precision in tunneling barrier thickness control. A critical issue in ALD of $Al_2O_3$ on metals is the lack of hydroxyl groups on metal surface, which prevents nucleation of the trimethylaluminum (TMA). In this work, we explore modifications of the aluminum surface with water pulse exposures followed by TMA pulse exposures to assess the feasibility of ALD as a viable technique for JJ qubits. ALD $Al_2O_3$ films from 40 Å to 100 Å were grown on 1.4 Å to 500 Å of Al and were characterized with ellipsometry and atomic force microscopy. A growth rate of 1.2 Å/cycle was measured, and an interfacial layer (IL) was observed. Since the IL thickness depends on the availability of Al and saturated at 2 nm, choosing ultrathin Al wetting layers may lead to ultrathin ALD $Al_2O_3$ tunneling barriers.**

## I. INTRODUCTION

Pristine dielectric films of thickness ~1-2 nm are required for many important applications including Josephson Junctions (JJs) that are building blocks for quantum bits (qubits) in quantum computers [1]. A JJ is typically fabricated on a superconductor-insulator-superconductor (S-I-S) trilayer, in which the thickness of the "I" layer is restricted to about ~1 nm to yield the desired supercurrent density, $J_c$. For "S" layers, aluminum or niobium are commonly employed and thermally oxidized aluminum ($AlO_x$) has been used as the tunneling

Manuscript received October 9, 2012. This work was supported by ARO contract W911NF-09-1-0295, W911NF-12-1-0412. JW also acknowledges support from NSF contracts NSF-DMR-1105986 and NSF EPSCoR-0903806, and matching support from the State of Kansas through Kansas Technology Enterprise Corporation.

A.J. Elliot, G. Malek, L. Wille, R.T. Lu, S.Y. Han, and J.Z. Wu are with the Department of Physics and Astronomy, University of Kansas, Lawrence, Kansas 66045, USA (corresponding author: A.J. Elliot, 785-864-2273; fax: 785-864-5262; e-mail: alane@ku.edu).

J. Talvacchio and R.M. Lewis, are with Northrop Grumman, Baltimore, MD 21203, USA (e-mail: john.talvacchio@ngc.com).

barrier. While $AlO_x$ can be readily generated on an Al electrode, a thin Al wetting layer of typical thickness in the range of 4-7 nm is incorporated on the Nb electrode for $AlO_x$ tunneling barrier fabrication.

The primary obstacle facing JJ qubits is the issue of decoherence. It is widely believed that microscopic defect states within the JJ circuit act as two level fluctuators (TLFs) and are the primary source of decoherence in JJ qubits[2]. These TLFs couple the JJs to the environment, destroying the entangled state of the qubit. Recently, improved coherence times have been reported which approach the fault tolerant computing limit by shielding the JJ and decoupling it from the environment[3]. An alternative and complementary approach to the decoherence problem is removing the TLFs from the device. Although the material source of the TLFs is not entirely understood, the defects in the insulating materials of the circuit are one source, and perhaps the most important source[2]. In particular, it has been shown that reducing the defect density in the tunneling barrier reduces the density of TLFs, but so far this has required using high temperature epitaxial techniques[4].

An alternative and unexplored method of eliminating defects in the tunneling barrier is Atomic Layer Deposition (ALD). ALD is a chemical vapor based layer-by-layer deposition process [5]. A nearly ideal ALD process is ALD $Al_2O_3$, which works by exposing a heated substrate to alternating pulses of $H_2O$ and trimethylaluminum (TMA) separated by a flush of nitrogen gas to assure the two chemicals never meet in a gaseous state. Growth occurs via ligand exchange between $H_2O$ and TMA at the substrate surface and is described by the chemical reactions [5]

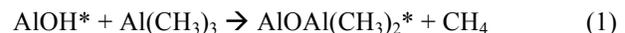

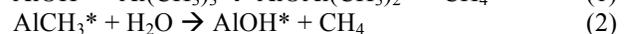

$$AlOH^* + Al(CH_3)_3 \rightarrow AlOAl(CH_3)_2^* + CH_4 \quad\quad (1)$$
$$AlCH_3^* + H_2O \rightarrow AlOH^* + CH_4 \quad\quad (2)$$

where an asterisk denotes a surface species. ALD is a relatively low temperature process with ALD $Al_2O_3$ typically occurring around 200°C. The chemical reactions involved only occur on the sample's surface, producing highly conformal growth. In each cycle of ALD, *i.e.* after both the reactions shown in Equations (1) and (2) have occurred once, only one molecular layer (ML) is produced, or about 1.1 Å of $Al_2O_3$ is coated [5]. This provides an atomic-scale control of film thickness. Furthermore, the sequential chemical reaction in the ALD process most likely results in a complete oxidation of every molecular layer of $Al_2O_3$, which differs from the physical diffusion in the thermal oxidation process and could lead to much lower defect densities. While the effective thickness of the tunneling barrier can be empirically controlled on the sub-Angstrom level using thermal oxidation[6], the benefits of increasing the coherence time of JJ qubits by reducing defects such as oxygen vacancies in the tunneling barrier may outweigh the cost of slightly decreased control of the barrier thickness. Therefore, ALD may improve the quality of the tunneling barrier in JJs, which in turn leads to improved coherence in JJ-based qubits. However, in order to test this prediction, a better understanding of ALD nucleation on Al substrates is necessary.

Though there are no previous reports of ALD $Al_2O_3$ on Al specifically, there are similar studies from which we may draw insight. In particular, extensive reports in the literature are available on ALD $Al_2O_3$ growth on Si due to its importance in semiconductor microelectronics. Theoretically, Si surfaces are terminated with a layer of hydroxyl groups, which allows TMA to attach readily in a similar way shown in Equation (1). However, an interfacial layer (IL) between $Al_2O_3$ and Si substrates has been reported which is typically ~1 nm thick and is composed of an alumina silicate [7]. Interestingly, a long exposure of Si to TMA can reduce the thickness of the IL to ~0.5 nm, while a long exposure to water does not have the same effect [8]. This suggests that the IL is caused in part by the thermal oxidation of the substrate by exposing it to water at elevated temperatures.

The metal surface may be much more complicated depending on the reactivity of the metal with oxygen, hydrogen and other species. In particular, the nucleation mechanism of ALD dielectric layers on metal surfaces is not well understood, though many efforts have been made in ALD growth of dielectric films on a wide variety of metallic substrates. Groner *et al.* [7] reported ALD growth of $Al_2O_3$ on Au, Co, Cr, Cu, Mo, Ni, NiFe, NiMn, Pt, PtMn, W, and even stainless steel. Only the $Al_2O_3$ films grown on Mo, Pt, and Au were selected for detailed electrical analysis. On Mo, the dielectric constant of the $Al_2O_3$ was within expected ranges, and an IL comparable to that found on Si substrates was observed. In contrast, the $Al_2O_3$ films grown on Pt and Au showed a reduced dielectric constant and growth rate (by about 15%) on Au as compared to the on Si case. This suggests that the nucleation of the ALD dielectric films like $Al_2O_3$ is more difficult on noble metal films. Several groups have studied ALD on noble metals in a greater detail [9],[10],[11],[12] and found the dielectric growth does not initiate during the first few ALD cycles. Instead, an incubation period is required to facilitate the nucleation of the ALD dielectric films on noble metals including Pt, Ir, and Ru. This incubation period can be avoided by directly hydroxylating the surface with a hydrous plasma [11], which suggests the nucleation difficulties spur from the thermodynamic unfavorability of noble metal oxides. To contrast this, other groups have studied ALD growth on other, more easily oxidized metals such as W, Co, and Ta [12][13][14]. It was found that the growth rate during the first couple of ALD cycles was greater than expected, and an IL formed between the substrate and the ALD film. This implies not only that ALD nucleation occurs easily when the substrate metal is easily oxidized, but also that an interfacial layer of thermal oxide is being formed. In particular, the thickness of the IL may be much larger than a few monolayers. For ultrathin dielectric layers on metal surfaces required for JJ qubits, it is therefore imperative and critical to understand the nucleation of dielectric ALD films in order to achieve the control of the morphology, thickness and, most importantly, defect density at the sub-nanometer scale.

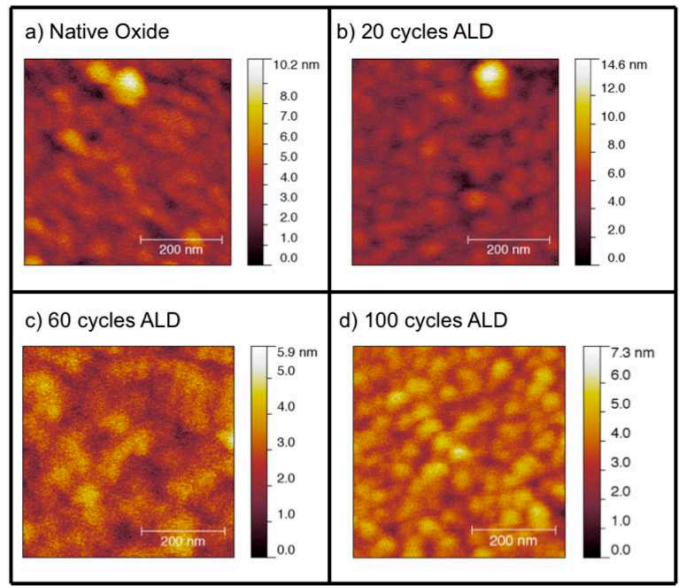

**Fig. 1:** Atomic force microscopy of a) the native aluminum oxide of 50 nm sputtered Al, b-d) 20, 60 and 100 cycles ALD on sputtered Al, respectively. The $R_{RMS}$ are 1.1 nm, 1.3 nm, 0.7 nm, and 0.8 nm respectively.

Al is a metal that can be readily oxidized even at high vacuum with traces of oxygen. This means it is plausible to assume a thin layer of $AlO_x$ may exist on the surface of the Al before ALD $Al_2O_3$ initiates. Nucleation of the TMA on the Al surface is anticipated even under *in situ* situations. The challenge in growth of ultrathin tunneling barriers on Al is therefore to minimize the thickness of this natural $AlO_x$ layer while using it to facilitate ALD $Al_2O_3$ nucleation. There are several possible ways to minimize the thickness of this IL, including ion milling to remove the formed IL, or by reducing the temperature, oxygen partial pressure, or heating time to frustrate its formation. However, the most straight forward and cost effective method, without introducing additional variables, is to simply vary the thickness of Al which is available to oxidize. In this work, ALD was performed *in situ* on sputtered Al surfaces. With variation of the Al film thickness, the focus of this work is on understanding the nucleation of $Al_2O_3$ on Al and developing schemes to obtain ultrathin ALD $Al_2O_3$ tunneling barrier for trilayers of Al-$Al_2O_3$-Al and Nb-$Al_2O_3$-Nb JJs for qubits.

## II. EXPERIMENTAL

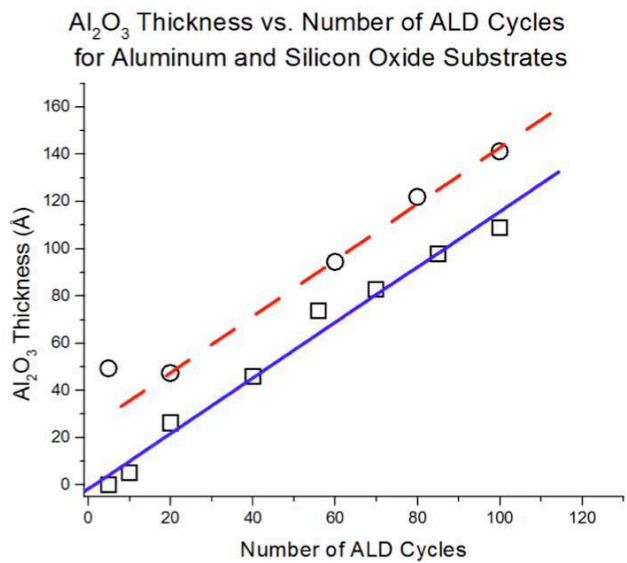

Fig. 2: Al₂O₃ thickness vs. ALD cycles for aluminum (red dashed, circles) and SiO₂ (blue solid, squares) substrates. The growth rate on both substrates is 1.2 Å/cycle.

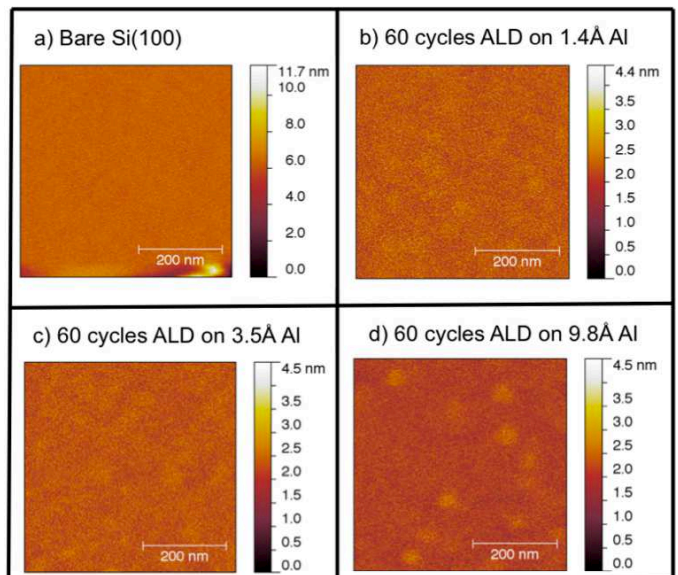

**Fig. 3:** Atomic force microscopy of a) Si(100), b-d) 60 cycles ALD on 1.4 Å, 3.5Å, and 9.8 Å Al, respectively. $R_{RMS}$ are are 0.6nm, 0.5nm, 0.5nm, and 0.4nm respectively.

To fabricate ultrathin films while minimizing the effects of ambient oxidation, a custom vacuum chamber was designed to integrate high vacuum sputtering with a home-made viscous flow ALD reactor[15, 16]. Al films were sputtered on SiO₂(500 nm)/Si and Si(100) substrates at either 15 W or 90 W DC from a 3" sputtering gun (Torus 3C from Kurt J. Lesker) in 14 milliTorr (mTorr) Ar. The sputtering rate was 0.07 nm/s for the former and 0.47 nm/s for the latter. The sample was then transported *in situ* to the ALD reaction chamber with a base pressure of approximately $10^{-6}$ Torr The ALD chamber was then heated to 200°C using resistive heaters under 500 mTorr from a 5 sccm N₂ flow over the course of about 1.5 hrs. ALD was then performed on the sample by exposing it to alternating pulses of H₂O and TMA. ALD was performed on 50 nm *in situ* sputtered Al, thermal SiO₂, and Si(100) with an *in situ* sputtered 1.4 Å to 9.8 Å Al wetting layer . The Si(100) substrates were prepared by stripping the SiO₂ with hydrofluoric acid just prior to loading the sample into the vacuum chamber. The surface morphology was studied with an atomic force microscope (AFM) from the WiTec company. The thicknesses of the Al₂O₃ films were measured with a Horiba UVISEL spectroscopic ellipsometer (SE) between 2.75 eV and 4 eV.

## III. RESULTS AND DISCUSSION

**Fig. 1** shows the surface morphology of the ALD Al₂O₃ films of various thicknesses grown on 50 nm thick sputtered Al measured using ex-situ AFM. The root-mean-square roughness ($R_{rms}$) was measured over a 5$\mu$m x 5$\mu$m scan area. **Fig. 1a)** shows the Al substrate with no ALD growth. The native oxide on Al, formed when the sample was exposed to air, has an $R_{rms}$ of 1.1 nm. Ripples can be seen with a lateral dimension of ~40 nm, a longitudinal dimension of ~100 nm, and height of ~5-10 nm. These ripples are randomly oriented and homogenously distributed across the surface. **Fig. 1(b-d)**, show Al₂O₃ films obtained with 20, 60, and 100 cycles of ALD growth, on which an $R_{rms}$ of 1.3 nm, 0.7 nm, and 0.8 nm were observed, respectively. The surface features on all samples are similar in size and shape. $R_{rms}$ was measured over other scan windows, and comparable results were obtained. These data show there is no correlation between surface morphology and Al₂O₃ thickness in this regime. The comparable $R_{rms}$ values across all samples and reappearance of surface ripples is confirmation that ALD Al₂O₃ growth is conformal on Al surfaces.

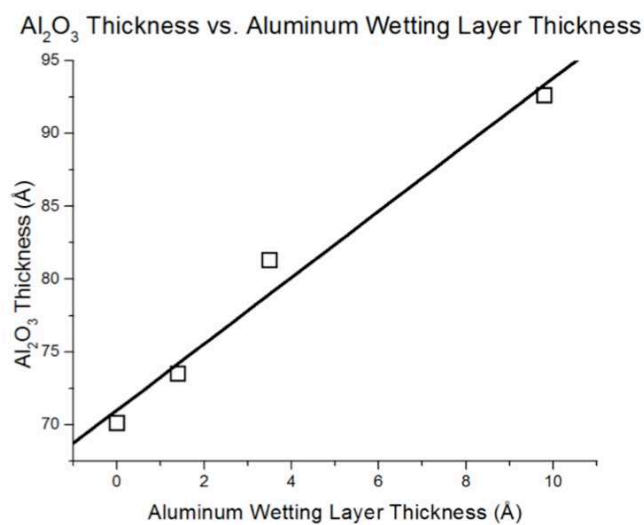

**Fig. 4: Al₂O₃ thickness vs Al wetting layer thickness. The Al was sputtered onto Si(100) and then 60 cycles of ALD were performed on all samples. The slope of the trendline , is 2.2.**

**Fig. 2** shows the results of the SE study of the ALD Al₂O₃ films of various thicknesses grown on SiO₂(500 nm)/Si and Al(50nm)/SiO₂/Si substrates. The blue solid line shows the ALD Al₂O₃ growth rate on the SiO₂ surface is 1.2 Å/cycle calculated from the slope of the curve, which is comparable to previously reported values [15]. The scatter in the data is a result of poor optical contrast between $Al_2O_3$ and $SiO_2$ due to their similar refractive indices. For sputtered Al substrates, exposing the film to ambient conditions for a few days formed a native oxide. This native oxide was measured to be 49 Å, and this set the lower limit for measuring ALD $Al_2O_3$ on Al due to oxygen's ability to diffuse through thin oxide layers. The red dashed line in **Fig. 2** shows the ALD $Al_2O_3$ growth rate on Al is also 1.2 Å/cycle, shown clearly in the figure since the two curves are parallel to each other. Thus, the growth rate of ALD $Al_2O_3$ is independent of the substrate after nucleation for the case of $SiO_2$ and Al, which differs from other works on noble metals [10]. However, there is significantly more growth of $Al_2O_3$ on Al than there is on $SiO_2$, which can be extracted by extrapolating the curve towards zero number of cycles. In fact, there is a systematic ~2 nm offset between the two curves. Since an equal growth rate was observed for ALD $Al_2O_3$ on Al and $SiO_2$ surfaces, this additional 2 nm of oxide on Al may be attributed to oxygen diffusion into Al most probably during sample heating before ALD fabrication process. The oxidation of the substrate after the ALD fabrication may be ruled out as the cause of this IL since 20 cycles of ALD produced an $Al_2O_3$ film of comparable thickness to the native oxide. Further, since the thickness of the IL is less than half of the thickness of the native oxide, the difference may be attributed to the lower partial pressure of oxygen in the vacuum chamber for ALD as compared to the ambient atmosphere. In addition, thinner IL was observed on SIS trilayers with fewer number of ALD cycles[16], suggesting the IL thickness correlates to the ALD growth time at 200 °C consistent to the expectation of the thermal oxidation process [6].

To explore ways to reduce the thickness of this IL to much below the thickness of the tunnel barrier, a wetting layer of ultrathin Al was sputtered onto Si(100) substrates before ALD growth. **Fig. 3** shows the surface morphology of a bare Si(100) substrate (**a**) and 60 cycles ALD $Al_2O_3$ on a Si(100) with ultrathin Al wetting layers of approximately 1.4 Å, 3.5 Å, and 9.8 Å (**b-d**), respectively. The thickness of the wetting layers was not directly measured. Instead, it was approximated using a previously calibrated sputtering rate. The $R_{rms}$ values over 5 $\mu$m x 5 $\mu$m are 0.6 nm, 0.5 nm, 0.5 nm, and 0.4 nm, respectively. There is no apparent trend between the morphology and the wetting layer thickness. Furthermore, comparing Fig. 1 to Fig. 2, the $R_{rms}$ values of ALD $Al_2O_3$ on ultrathin Al wetting layers is significantly smaller than on 50 nm Al. This is because $R_{rms}$ for Si(100) is significantly smaller than $R_{rms}$ for 50 nm Al and because ALD growth is conformal. **Fig. 4** shows the $Al_2O_3$ thickness vs. wetting layer thickness. Interestingly, the $Al_2O_3$ thickness decreases monotonically with the Al wetting layer thickness, confirming that the IL is indeed formed via oxidation of the Al surface layer. In particular, the IL could be almost removed and the surface roughness could be improved by using an extremely thin Al wetting layer.

A preliminary proof of concept study on tunnel junctions with ALD tunneling barriers, such as Nb-$Al_2O_3$-Nb with Al wetting layers of 7 nm thickness, have shown promising results and have been published elsewhere[16]. This wetting layer thickness is typical in JJ fabrication with thermally oxidized tunneling barriers and was chosen to allow direct comparison between these and ALD tunneling barriers. The tunneling properties of these trilayers were characterized at room temperature using the Current-In-Plane-Tunneling (CIPT) technique, which confirmed the anticipated tunneling barrier resistance as a function of the ALD cycles and, thus, barrier thickness. CIPT also revealed excellent uniformity across 2×2 $cm^2$. JJs were fabricated from these trilayers and low temperature I-V curves were obtained. Preliminary results suggest very low subgap leakage current in the ALD tunneling barrier with 8 cycles of ALD. The ALD junction has a high specific resistance $R_N A = 3.57$ k$\Omega\mu m^2$, which is more than two orders of magnitude larger than that of the reference sample without exposure to any ALD source vapor, where $R_N$ is normal state resistance and A is junction area. We speculate that this significant increase in $R_N$ could possibly be caused in part by the IL described in this work. Using the empirical result of $eI_cR_N/\Delta = 1.27$ previously measured in Nb/AlO$_x$/Nb junctions in our lab[17], where $\Delta$ and $e$ are the superconducting gap energy and the charge of an electron, respectively, a critical current density of about 39 A/$cm^2$ at 4.2 K can be obtained for the ALD junction[16]. The measured $I_cR_N$ product is smaller than 0.3 mV for the ALD junction, indicating the Ic was suppressed and might be caused by the OH termination of the ALD $Al_2O_3$ layer, which could act as a scattering center for cooper pairs, suppressing the supercurrent to near zero. Further work exploring the effects of using ultrathin Al wetting layers and optimizing the junction performances are currently underway.

## IV. CONCLUSIONS

In conclusion, these results show that self-limited, conformal ALD $Al_2O_3$ growth could be readily achieved on Al. The growth rate of ALD $Al_2O_3$ on Al and $SiO_2$ substrates is 1.2 Å/cycle, which agrees well with previously reported values of ALD $Al_2O_3$ on a variety of other substrates. However, ILs of ~2 nm were formed on thick Al wetting layers, which may prevent achievement of ultrathin $Al_2O_3$ tunneling barriers of thickness on the order of 1-2 nm. This issue could be resolved by adopting ultrathin Al wetting layers, and we have demonstrated an Al wetting layer as thin as 1.4 Å is viable. This suggests that after nucleation the ALD $Al_2O_3$ is indeed self-limited, but the nucleation on metal surfaces requires interfacial engineering to minimize the unwanted ILs driven by oxygen diffusion. Further, the surface roughness of the $Al_2O_3$ films is comparable to the substrate upon which they were grown, proving conformity and self-limitation. On ultrathin Al, the IL thickness is a function of Al availability. Tighter control of the IL thickness must be achieved to create ultrathin tunneling barriers for Josephson junction based qubits.

## Acknowledgements

This research was supported by ARO contract W911NF-09-1-0295, W911NF-12-1-0412. JW also acknowledges support from NSF contracts NSF-DMR-1105986 and NSF EPSCoR-0903806, and matching support from the State of Kansas through Kansas Technology Enterprise Corporation.

[1] G. Wendin and V. S. Shumeiko, "Quantum bits with Josephson Junctions," *Low Temperature Physics,* vol. 33, pp. 724-744, 2007.

[2] R. McDermott, "Materials Origins of Decoherence in Superconducting Qubits," *IEEE Transactions on Applied Superconductivity,* vol. 19, pp. 2-13, 2009.

[3] C. Rigetti, J. M. Gambetta, S. Poletto, B. L. T. Plourde, J. M. Chow, A. D. Córcoles, J. A. Smolin, S. T. Merkel, J. R. Rozen, G. A. Keefe, M. B. Rothwell, M. B. Ketchen, and M. Steffen, "Superconducting qubit in a waveguide cavity with a coherence time approaching 0.1 ms," *Physical Review B,* vol. 86, p. 100506, 2012.

[4] S. Oh, K. Cicak, J. S. Kline, M. A. Sillanpää, K. D. Osborn, J. D. Whittaker, R. W. Simmonds, and D. P. Pappas, "Elimination of two level fluctuators in superconducting quantum bits by an epitaxial tunnel barrier," *Physical Review B,* vol. 74, p. 100502, 2006.

[5] S. M. George, "Atomic Layer Deposition: An Overview," *Chemical Review,* vol. 110, pp. 111-131, 2010.

[6] N. Cai and G. Zhou, "Tuning the Limiting Thickness of a Thin Oxide Layer on Al(111) with Oxygen Gas Pressure," *Physical Review Letters,* vol. 107, p. 035502, 2011.

[7] M. D. Groner, J. W. Elam, F. H. Fabreguette, and S. M. George, "Electrical characterization of thin Al2O3 films grown by atomic layer deposition on silicon and various metal substrates," *Thin Solid Films,* vol. 413, pp. 186-197, 2002.

[8] M. Xu, H. Lu, S. Ding, L. Sun, W. Zhang, and L. Wang, "Effect of Trimethyl Aluminium Surface Pretreatment on Atomic Layer Deposition Al2O3 Ultra-Thin Film on Si Substrate," *Chinese Physics Letters,* vol. 22, p. 2418, 2005.

[9] K. Kukli, T. Aaltonen, J. Aarik, J. Lu, M. Ritala, S. Ferrari, A. Hårsta, and M. Leskelä, "Atomic Layer Deposition and Characterization of HfO2 Films on Noble Metal Film Substrates," *Journal of the Electrochemical Society,* vol. 152, pp. F75-F82, 2005.

[10] K. Kukli, M. Ritala, T. Pilvi, T. Aaltonen, J. Aarik, M. Lautala, and M. Leskelä, "Atomic layer deposition rate, phase composition and performance of HfO2 films on noble metal and alkoxylated silicon substrates," *Materials Science and Engineering B,* vol. 118, pp. 112-116, 2005.

[11] C. Chang, Y. Chiou, C. Hsu, and T. Wu, "Hydrous-Plasma Treatment of Pt Electrodes for Atomic Layer Deposition of Ultrathin High- k Oxide Films," *Electrochemical Solid State Letters,* vol. 10, pp. G5-G7, 2007.

[12] C. L. Platt, N. Li, K. Li, and T. M. Klein, "Atomic layer deposition of HfO2: Growth initiation study on metallic underlayers," *Thin Solid Films,* vol. 518, pp. 4081-4086, 2010.

[13] R. K. Grubbs, C. E. Nelson, N. J. Steinmetz, and S. M. George, "Nucleation and growth during the atomic layer deposition of W on Al2O3 and Al2O3 on W," *Thin Solid Films,* vol. 467, pp. 16-27, 2004.

[14] S. Y. Lee, H. Kim, P. C. McIntyre, K. C. Saraswat, and J. Byun, "Atomic layer deposition of ZrO2 on W for metal–insulator–metal capacitor application," *Applies Physics Letters,* vol. 82, pp. 2874-2876, 2003.

[15] J. W. Elam, M. D. Groner, and S. M. George, "Viscous flow reactor with quartz crystal microbalance for thin film growth by atomic layer deposition," *Review of Scientific Instruments,* vol. 73, pp. 2981-2987, 2002.

[16] R. Lu, A. J. Elliot, L. Wille, B. Mao, S. Han, J. Z. Wu, J. Talvacchio, H. M. Schulze, R. M. Lewis, D. J. Ewing, H. F. Yu, G. M. Xue, and S. P. Zhao, "Fabrication of Nb/Al2O3/Nb Josephson Junctions Using In Situ Magnetron Sputtering and Atomic Layer Deposition," *Applied Superconductivity, IEEE Transactions on,* vol. 23, pp. 1100705-1100705, 2013.

[17] H. F. Yu, X. B. Zhu, Z. H. Peng, Y. Tian, D. J. Cui, G. H. Chen, D. N. Zheng, X. N. Jing, L. Lu, S. P. Zhao, and S. Han, "Quantum Phase Diffusion in a Small Underdamped Josephson Junction," *Physical Review Letters,* vol. 107, p. 067004, 2011.